\documentclass[longbibliography,prb,preprint]{revtex4-1}

\usepackage{amsmath}
\usepackage{array}          
\usepackage{enumerate}
\usepackage{graphicx}
\usepackage[font=itshape]{quoting}          
\usepackage{siunitx}
\usepackage{tabularx}
\usepackage[normalem]{ulem}             
\usepackage{filecontents}               

\newcommand{\question}[2]{
    \begin{tabularx}{\textwidth}{@{}l X@{}}
        \textbf{#1} & #2
    \end{tabularx}
}                                                  

\renewcommand{\marks}[1]{\vspace{-0.5\baselineskip}\begin{flushright}$[#1]$\end{flushright}}

\newcolumntype{C}[1]{>{\centering\arraybackslash}m{#1}}        

\begin{document}

\title{Analyzing student conceptual understanding of resistor networks using binary, descriptive, and computational questions}

\author{Abid H. Mujtaba}
\email{abid.mujtaba@comsats.edu.pk}
\affiliation{Department of Physics, COMSATS Institute of Information Technology, Park Road, Tarlai Kalan, Islamabad 45550, Pakistan}

\date{\today}

\begin{abstract}
    This paper presents a case-study assessing and analyzing student engagement with and responses to binary, descriptive, and computational questions testing the concepts underlying resistor networks (series and parallel combinations). The participants of the study were undergraduate students enrolled in a university in Pakistan. The majority of students struggled with the descriptive question, even when successfully answering the binary and computational ones, failed to build an expectation for the answer, and betrayed significant lack of conceptual understanding in the process. The data collected was also used to analyze the relative efficacy of the three questions as means of assessing conceptual understanding. The three questions were revealed to be uncorrelated and unlikely to be testing the same construct. The ability to answer the binary or computational question was observed to be divorced from a deeper understanding of the concepts involved.
\end{abstract}

\maketitle

\section{Introduction}

    Assessment forms an integral part of education and Physicists are increasingly turning to data gathering and analysis to refine this crucial component of pedagogy.\cite{wieman_sci_method, wieman_transforming, von2016secondary, williamson2016applicability} The purpose of assessment in a physics course is to ostensibly determine the degree of conceptual understanding achieved by each student. Ostensibly, since inertia often leads both instructors and students to consider assessment as a proforma part of any course. It is therefore essential that instructors continuously evaluate their assessment techniques to ensure that these continue to provide valid insight in to student understanding.

    While a great deal of work has been carried out assessing and analyzing students studying physics in the developed world\cite{von2016secondary, williamson2016applicability} corresponding research in the developing world\cite{hussain2011physics} remains sparse. This paper attempts to begin to fill this gap as far as physics pedagogy in Pakistan is concerned by presenting a case-study that provided insight in to how a certain group of Pakistani students engaged with three types of questions (binary, computational, and descriptive) based on a single physics concept (combining resistors). The data obtained also allowed a comparison of the relative efficacy of these three types of questions as a means of judging student conceptual understanding.

\section{Case-Study}

    The case-study consisted of a carefully constructed three-part question which was made part of a mid-term examination taken by a particular group of students.

    \subsection{Students}

        The students who attempted this question were enrolled at the COMSATS Institute of Information Technology (CIIT) in Islamabad in Fall 2016. They were registered in the Electronics program of the Department of Physics and were in the third or fourth semesters of their four-year undergraduate degree.

        28 students attempted the question of which 9 were female. Although specific details about their educational and socio-economic background were not acquired CIIT generally caters to students belonging to middle-class backgrounds. These students have diverse educational histories on account of the bifurcation in Pakistan's educational system which comprises of government run public schools and a parallel system of private schools catering to about 40\% of school-going children.\cite{dawn_private_schools} The quality of education imparted in these parallel systems and in the schools within each system is understood to be extremely inhomogeneous\cite{siddiqui2017comparing} with private schools widely considered to be a better alternative.\cite{aldermana2001school}

        The medium of instruction and examination at CIIT is English but the students are not native speakers of it. This constitutes a significant source of pedagogic friction with most classes conducted in a bilingual setting, continuously switching between English and Urdu.\cite{malik2010code} All examinations are, however, conducted entirely in English.

    \subsection{Course}

        The students in questions were registered in a course titled ``Circuit Theory'' (code PHY221) which is considered the first foundational course of the Electronics program. Instruction comprised of two 1.5 hour lectures and a single 3 hour lab per week in a 16-week long semester (excluding end-of-term exams). The primary textbook used was ``Basic Engineering Circuit Analysis''\cite{circuit_theory_book} by Irwin \& Nelms. The students had already taken two courses in Calculus and a pre-requisite physics course titled ``Electricity, Magnetism, and Optics''.

    \subsection{Concept Tested}

        The three-part question that formed the basis of the case-study was based on the topic of resistor networks, there being a rich history of physics education research using this topic.\cite{newman2017first,leniz2017students, engelhardt2004students, etkina2006using} The concept being tested was series and parallel combinations of resistors, specifically the understanding that series resistance is greater than the sum of the parts and parallel resistance is less than all constituent parts.

    \subsection{The Question}

        The case-study revolved around the third question asked in an \textbf{open-book} exam consisting of three questions. The entire exam was 80 minutes long with the expectation that students would spend at least 20 minutes on this question. The question consisted of three parts with the first two scored out of $2$ and the last part out of $4$.

        \vspace{-0.5\baselineskip}
        \begin{center}\rule{\linewidth}{0.5pt}\end{center}

\begin{figure}[ht!]
                \includegraphics{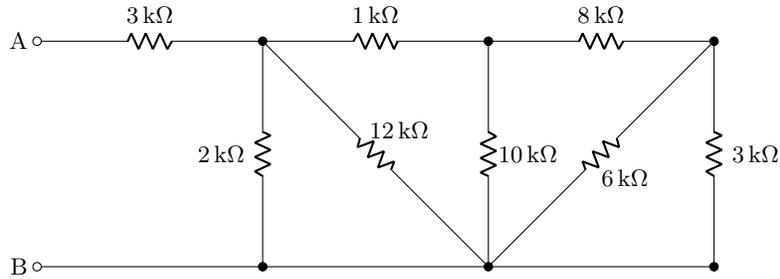}
                \caption{\label{resistor_network} Resistor network}
            \end{figure}

        \question{Q.3}{Consider the following resistor network (in Fig.~\ref{resistor_network}).}

            \vspace{\baselineskip}
            Solve the first two parts \textbf{without} calculating the equivalent/net resistance $R_{AB}$ of the network.

            \begin{enumerate}[(i)]
                \item Is $R_{AB} > \SI{3}{\kilo\ohm} \,$? Justify your answer.
                \item Is $R_{AB} > \SI{5}{\kilo\ohm} \,$? Justify your answer.
                \item Calculate $R_{AB}$.
            \end{enumerate}

            \marks{2 + 2 + 4}
            \vspace{-1.5\baselineskip}

        \begin{center}\rule{\linewidth}{0.5pt}\end{center}

    \subsection{Types of Questions}

        Each response was divided in to three parts for evaluation based on the three types of questions that constituted the whole.

        \subsubsection{Binary}

            The first half of parts (i) and (ii) were of the \textbf{binary} type a variant of the limited response type (the most common example of which is the MCQ). The only valid response to these was either Yes or No.

            The students were instructed to come up with an answer before performing a detailed calculation. It was hoped that this would inspire the students to think about the problem logically and apply their conceptual understanding of series and parallel resistances to come up with the answers.

            The students could then use these inequalities ($3 < R_{AB} < 5$) to form an expectation for the answer to part (iii) allowing them to detect mistakes. Continuous expectation building and evaluation while solving problems is a crucial skill which needs to be imparted to the students.

        \subsubsection{Descriptive}

            The second half of parts (i) and (ii) required the students to provide justification (in the form of a few descriptive sentences) for their response to the first half. It was designed to probe a deeper understanding of the concept of series and parallel combinations of resistors, ascertaining if the students understood the abstract idea that resistance increases when added in series and reduces when added in parallel.

            It is the objective of any physics course to use lectures, assignments, quizzes, and exams to impart this deeper understanding of abstractions to students and the descriptive part was designed to assess this aspect.

        \subsubsection{Computational}

            Part (iii) required the students to carry out a relatively detailed calculation to arrive at an answer. It was intended to test students' problem solving skills as well as their conceptual understanding. These skills were considered to be the closest to what a practitioner would use in a real-world application.

\section{Evaluation}

    \subsection{Binary}

        The binary questions could only be answered with a Yes or No response. One point was given for each correct answer (Yes for (i) and No for (ii)).

    \subsection{Descriptive}

        Since resistances are added when in series it is clear from the \SI{3}{\kilo\ohm} resistor next to port A that the net resistance must be greater than \SI{3}{\kilo\ohm}. Similarly since the remainder of the circuit is in parallel with the \SI{2}{\kilo\ohm} resistor it follows that the net resistance for that part is less than \SI{2}{\kilo\ohm} and consequently the overall resistance is less than \SI{5}{\kilo\ohm}.

        Students were evaluated on how close they came to this reasoning. Reproduced below are a few representative responses, presented verbatim. Legibility was a concern, unfortunately, the students' hand-writing could not be reproduced digitally.

        \subsubsection{Above-average response}

        \begin{quoting}
            Yes $R_{AB} > 3$ because net resistance except \SI{3}{\kilo\ohm} is a number which we will add in \SI{3}{\kilo\ohm} to calculate equilant resistance of whole circuit that's why when any number is added in \SI{3}{\kilo\ohm} it will be greater the \SI{3}{\kilo\ohm}.

            No $R_{AB} > 5$ bcz net resistance of circuit except \SI{3}{\kilo\ohm} and \SI{2}{\kilo\ohm} is a number and \SI{3}{\kilo\ohm} and \SI{2}{\kilo\ohm} are in series when net resistance will be add in \SI{2}{\kilo\ohm} + net rest. then the sum is less than \SI{2}{\kilo\ohm} and then this sum is added in \SI{3}{\kilo\ohm} so it will not reach \SI{5}{\kilo\ohm} bcz it is less than \SI{2}{\kilo\ohm}.
        \end{quoting}

        Despite the grammatical and spelling mistakes it was quite clear that the student was applying concepts correctly. The exposition left much to be desired. A decision was made to grade liberally and ignore the quality of language and presentation. This response was awarded full (two) points.

        The communicative quality of this, one of the better responses, revealed a serious pedagogical issue. It is not enough for someone studying physics to understand a concept, it is necessary that she be able to communicate scientific thought in a clear and rigorous manner. The responses were lacking in this aspect, despite having access to textbooks that exhibit this very quality. The descriptive question underlined a lack in a core skill providing crucial feedback to the instructor.

        The fact that the students are not native speakers of English does play a part, but for exactly this reason the students need to be instructed to focus even more on communicating in the language of mathematics. The descriptive question can be answered with minimal recourse to English with an appropriate mix of figures and equations. The response should be treated as a mathematical proof. Not only is this an end in itself, striving for it forces students to think logically.

        \subsubsection{Average response}

        \begin{quoting}
            Yes $R_{AB}$ is greater than \SI{3}{\kilo\ohm} because $R_{AB}$ is equivalent of this whole circuit and there are many resistances in this circuit.

            $R_{AB}$ is not greater than \SI{5}{\kilo\ohm} because the sum of \SI{3}{\kilo\ohm} and \SI{2}{\kilo\ohm} is \SI{5}{\kilo\ohm}. So $R_{AB}$ should not be greater than \SI{5}{\kilo\ohm}.
        \end{quoting}

        This response received 0.5 points out of 2. The first statement is nearly incoherent while the second one skates close to the answer without ever arriving at it. Put together they betray significant conceptual weakness. However, this student managed to earn full points in the binary question and 75\% points in the computational question.

        \subsubsection{Below-average response}

        The underlined words are the ones added over the struck-out words.

        \begin{quoting}
            \sout{Yes} \uline{No}, $R_{AB}$ is \sout{greater} \uline{Not} than \SI{3}{\kilo\ohm} because the circuit shows series-parallel resistence and $R_{AB}$ is \sout{greater} \uline{Not} than \SI{3}{\kilo\ohm}.

            No, $R_{AB}$ is not greater than \SI{3}{\kilo\ohm}.
        \end{quoting}

        This student arrived at an answer of \SI{1.7}{\kilo\ohm} for the computational part (as compared to the correct answer of \SI{4.33}{\kilo\ohm}). The struck out parts suggest that the changes were made after this erroneous computational result was arrived upon. Rather than question the computational result on the basis of the reasoning used for the descriptive response the student chose to overwrite the response to fit with the arrived upon result. An unfortunate inversion of the expectation-based approach to problem-solving that was the purpose of this question. Nonetheless, a fascinating insight in to the thinking process of the student which would have been totally inaccessible if simply the binary and/or computational questions had been asked.

    \subsection{Computational}

        The computational question asked the student to calculate the net resistance of the circuit. This required the student to start from the right of the circuit and apply the equations for series and parallel resistors iteratively to arrive upon the final answer (\SI{4.33}{\kilo\ohm}).

        Quarter points were deducted if the student failed to write down the units at any stage where an interim result was stated. Half a point was deducted whenever the student made a calculation error (the most common one being errors while adding fractions). A full point was deducted whenever a student failed to identify a series or parallel combination correctly or made a mistake in simplifying the circuit after certain resistors had been combined.

\section{Data Analysis}

\begin{figure}[!ht]
        \begin{tabular}{|c|c|}
            \hline
            \parbox[c]{0.5\textwidth}{\vspace{0.5\baselineskip}\includegraphics[width=0.45\textwidth]{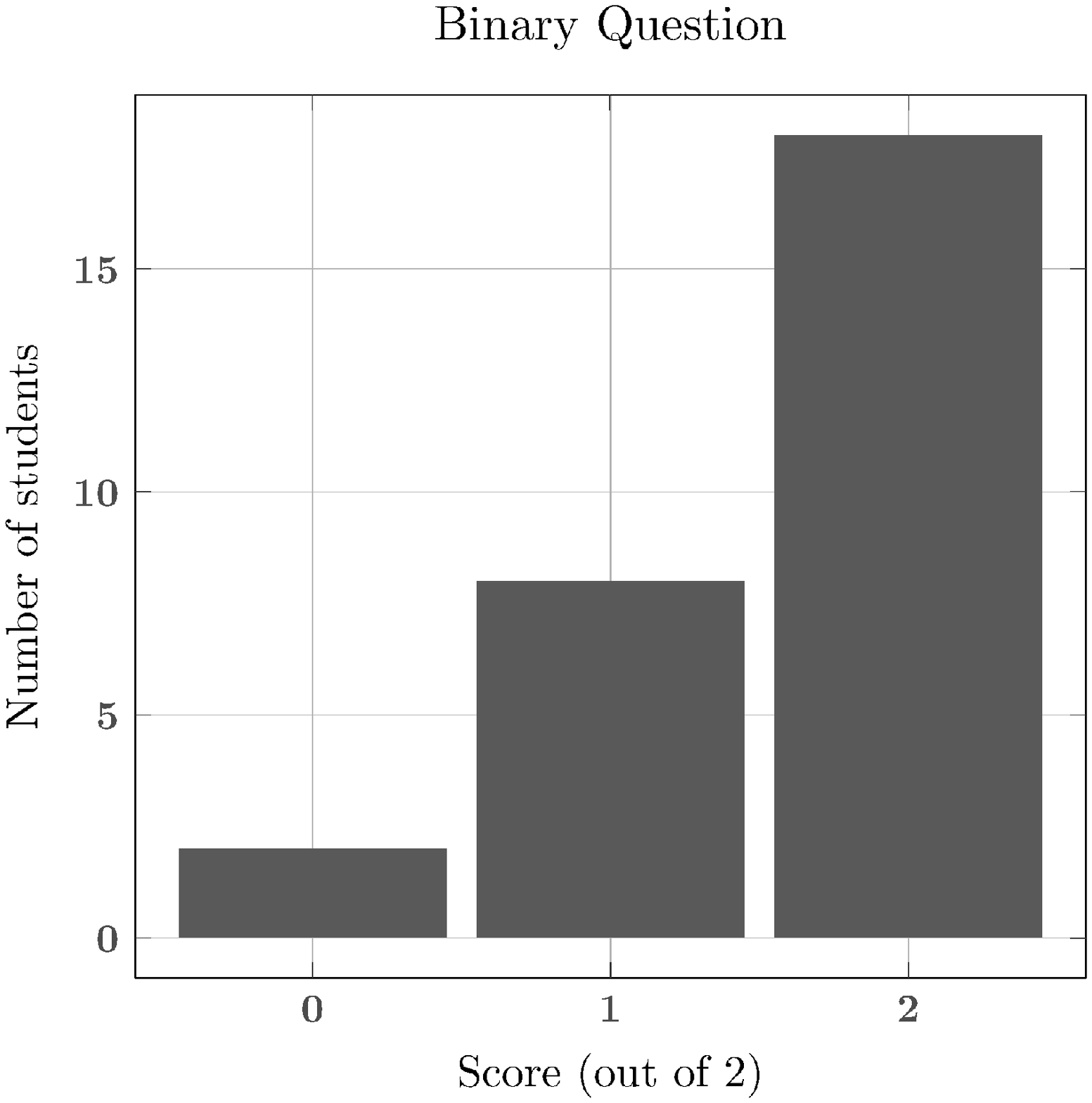}\vspace{0.5\baselineskip}} & \parbox[c]{0.5\textwidth}{\includegraphics[width=0.45\textwidth]{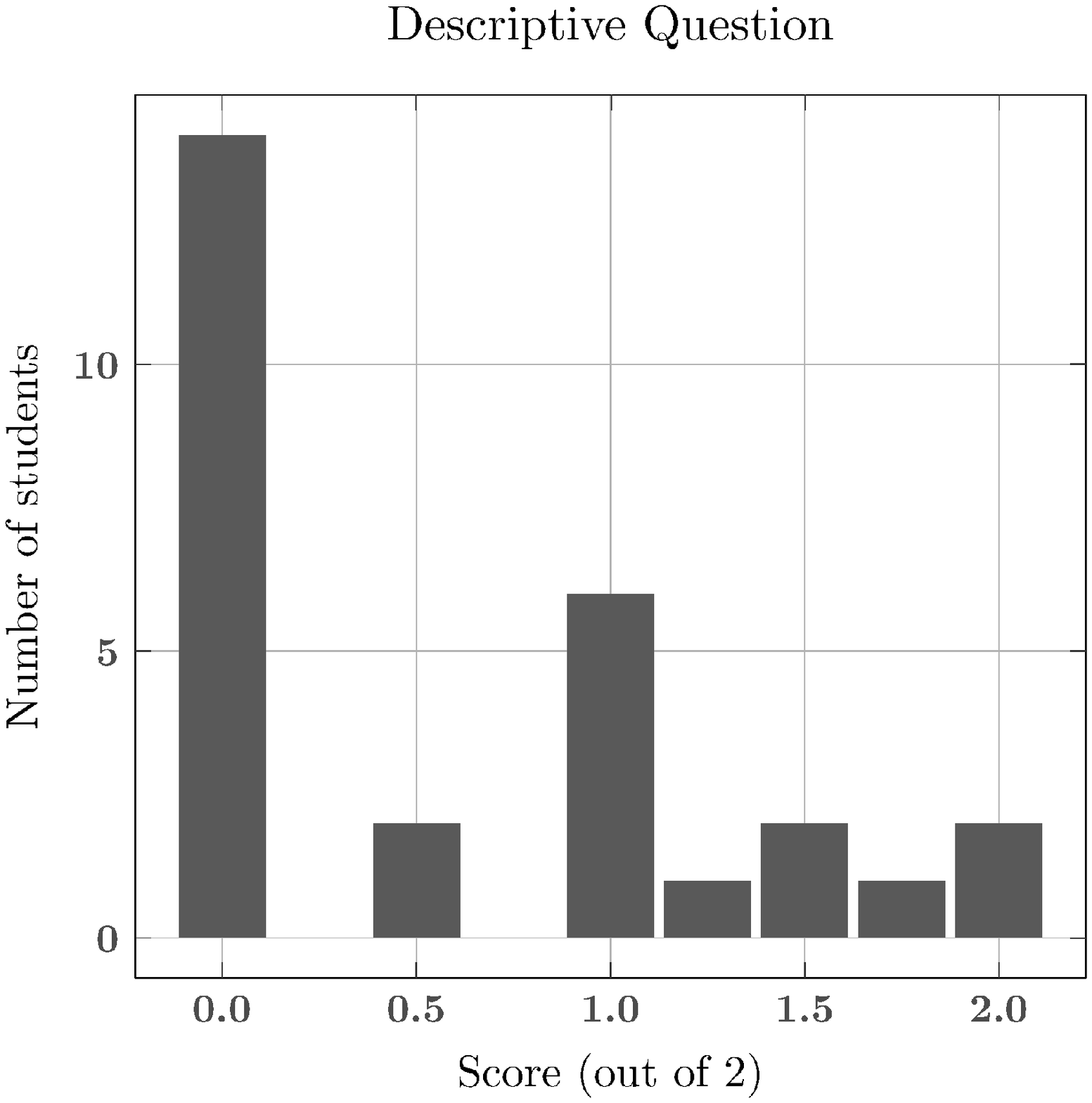}} \\
            \hline
            \parbox[c]{0.5\textwidth}{\vspace{0.5\baselineskip}\includegraphics[width=0.45\textwidth]{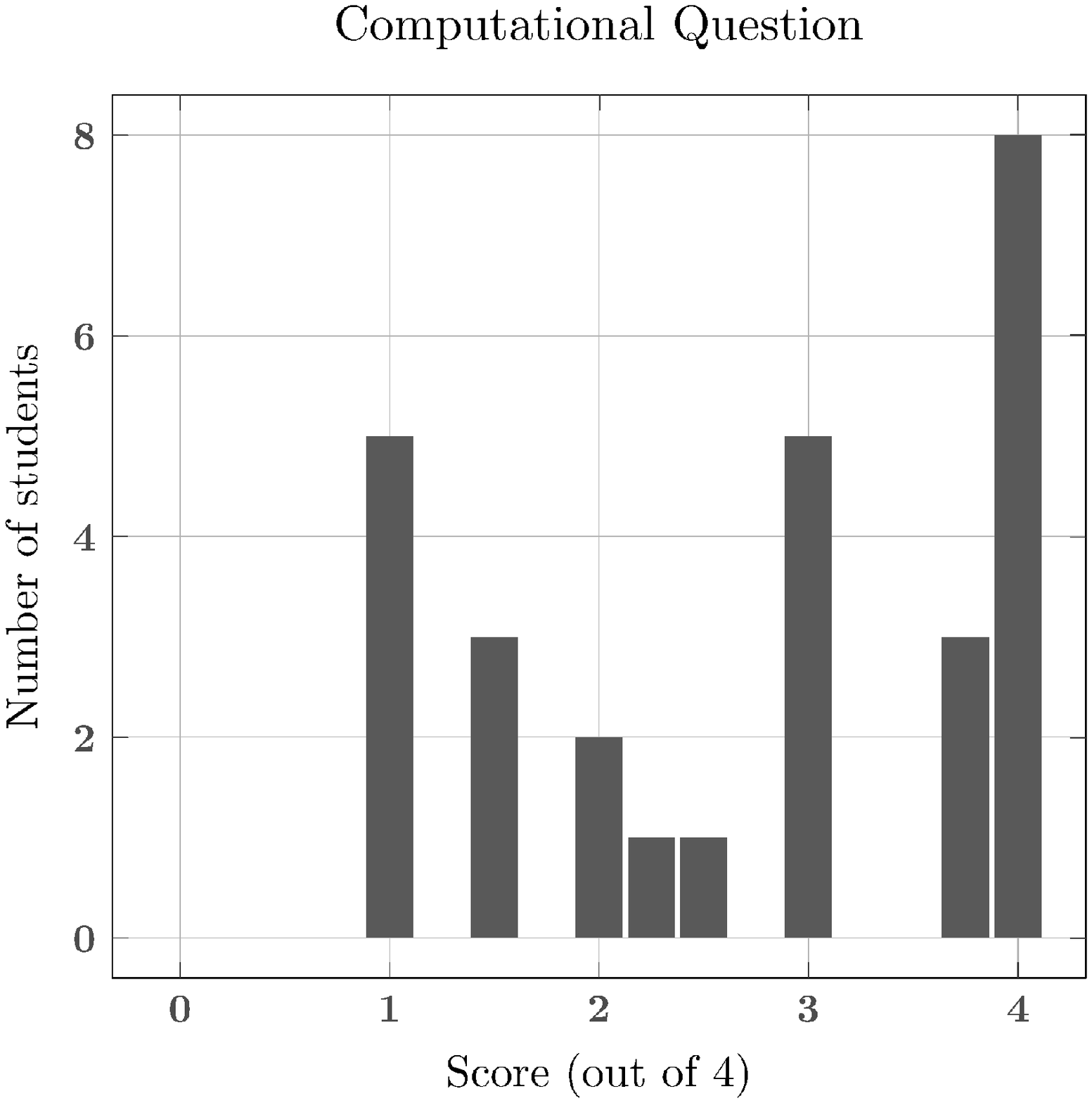}\vspace{0.5\baselineskip}} & \parbox[c]{0.5\textwidth}{\includegraphics[width=0.45\textwidth]{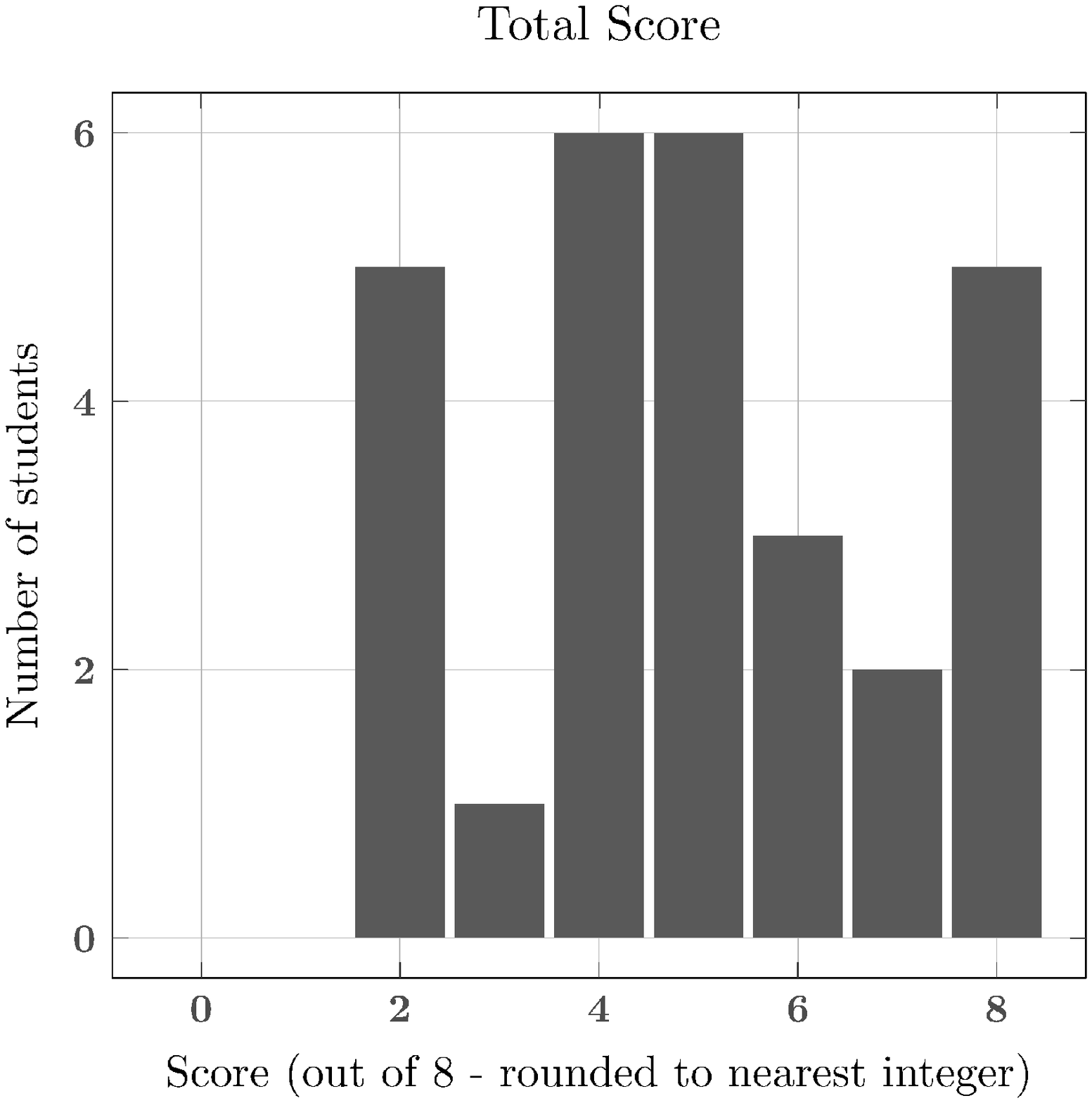}} \\
            \hline
        \end{tabular}
        \caption{\label{histograms} Histograms of the scores of the binary, descriptive and computational questions, and the total score.}
    \end{figure}

    The collected raw data, in the form of the points awarded to each type of question, has been made available online.\cite{raw_data}
    The mean score for each type of question along with the (population) standard deviation is given in Table~\ref{mean_and_sd} while Fig.~\ref{histograms} gives the distribution of scores for each type of question.

    \begin{table}[!ht]
        \begin{center}
            \caption{\label{mean_and_sd} Mean and (Population) Standard Deviation}
            \begin{ruledtabular}
                \begin{tabular}{ C{8em} *{3}{C{5em}} }
                    & max & mean & s.d. \\
                    \hline
                    binary & $2.00$ & $1.57$ & $0.62$ \\
                    descriptive & $2.00$ & $0.61$ & $0.69$ \\
                    computational & $4.00$ & $2.73$ & $1.15$ \\
                    total & $8.00$ & $4.91$ & $1.87$ \\
            \end{tabular}
            \end{ruledtabular}
        \end{center}
    \end{table}

    The data suggests that students performed the best in the binary question and worst in the descriptive question. The difference between the normalized mean values are significant. This can possibly be due to the clearly different levels of difficulty and gradation of the three questions, however, the score distribution across the student population differs radically between the three questions. While more than half of the students scored full points in the binary questions more than half of the students failed to score any points in the descriptive question.

    This raised concerns about the relative efficacy of these questions as metrics for judging conceptual competence which was the ultimate goal of the examination. The three questions were painting different pictures of the students' ability to reason about resistor networks.

\section{Relative Efficacy}

    A number of techniques were employed to compare the students' responses to the three types of questions in an attempt to determine their relative efficacy as a means of assessing student conceptual understanding of resistor networks. The insights gained may be generalizable to the assessment of a broader category of concepts by these types of quesitons.

    \subsection{Tau Equivalent Reliability ($\rho_T$)}

        Also known as Cronbach's $\alpha$, $\rho_T$ provides an estimate of the reliability of the assumption that a set of tests measure the same construct. $\rho_T$ does not measure the validity of the assumption, the tests could reliably be measuring some construct other than the assumed one, so $\rho_T$ is a necessary but not a sufficient condition\cite{tavakol2011making} meaning a negative result is more revealing than a positive one. The three questions in this study were assumed to be measuring understanding of the same concept (series and parallel resistors). If that was the case $\rho_T$ would reveal the reliability, or lack there of, of this assumption.

        The value of $\rho_T$ for the three paired combinations of questions and all three taken together\footnote{$\rho_T$ was calculated using open-source software R.\cite{r_site} The source code\cite{r_code} has been made available online} are given in Table~\ref{rho_T}.

        \begin{table}[!ht]
            \begin{center}
                \caption{\label{rho_T} Tau Equivalent Reliability}
                \begin{ruledtabular}
                    \begin{tabular}{ C{25em} C{20em} }
                        & $\rho_T$ \\
                        \hline
                        binary + descriptive & $0.61$ \\
                        descriptive + computational & $0.58$ \\
                        computational + binary & $0.21$ \\
                        all three & $0.56$ \\
                    \end{tabular}
                \end{ruledtabular}
            \end{center}
        \end{table}

        The closer the value of $\rho_T$ is equal to 1 the more likely it is that the combination of tests is measuring the same construct with values of $\rho_T > 0.7$ considered acceptable.\cite{wallace2010concept}

        Our calculations revealed that it was unlikely that these three types of questions were measuring the same construct. To dig a little deeper we looked at the correlation between the responses to these three questions.

    \subsection{Correlation Coefficient ($r$)}

        ``Pearson correlation coefficient'' in full, $r$ is an estimate of the degree of linear correlation between two variables. Table~\ref{cor_coeff} contains the correlation coefficients calculated between the responses to the three questions.

        \begin{table}[!ht]
            \begin{center}
                \caption{\label{cor_coeff} Correlation Coefficients}
                \begin{ruledtabular}
                    \begin{tabular}{ C{25em} C{20em} }
                        & $r$ \\
                        \hline
                        binary + descriptive & $0.44$ \\
                        descriptive + computational & $0.46$ \\
                        computational + binary & $0.14$ \\
                    \end{tabular}
                \end{ruledtabular}
            \end{center}
        \end{table}

\begin{figure}[!ht]
            \begin{tabular}{|c|c|}
                \hline
                \parbox[c]{0.5\textwidth}{\vspace{0.5\baselineskip}\includegraphics[width=0.45\textwidth]{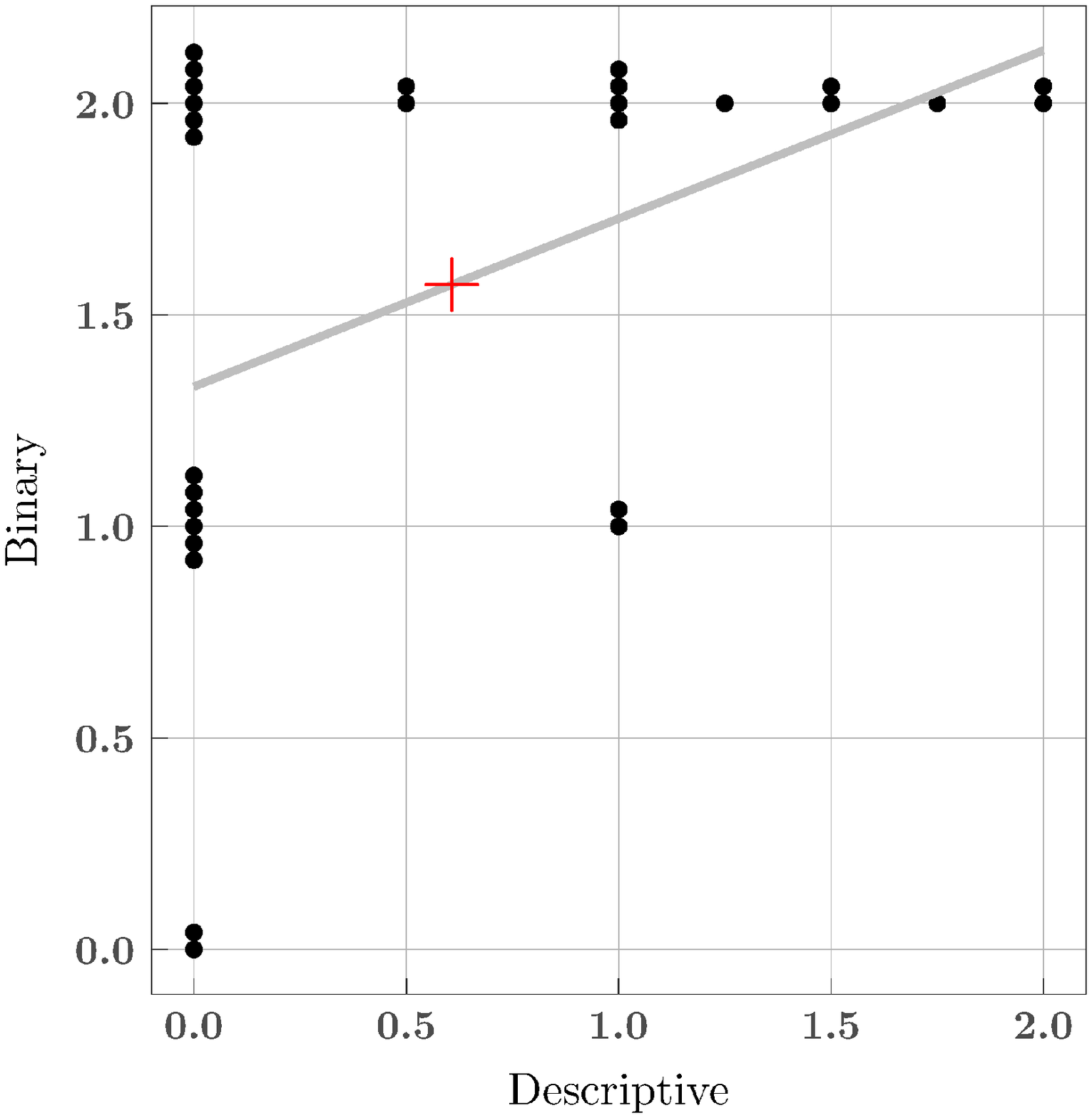}\vspace{0.5\baselineskip}} & \parbox[c]{0.5\textwidth}{\includegraphics[width=0.45\textwidth]{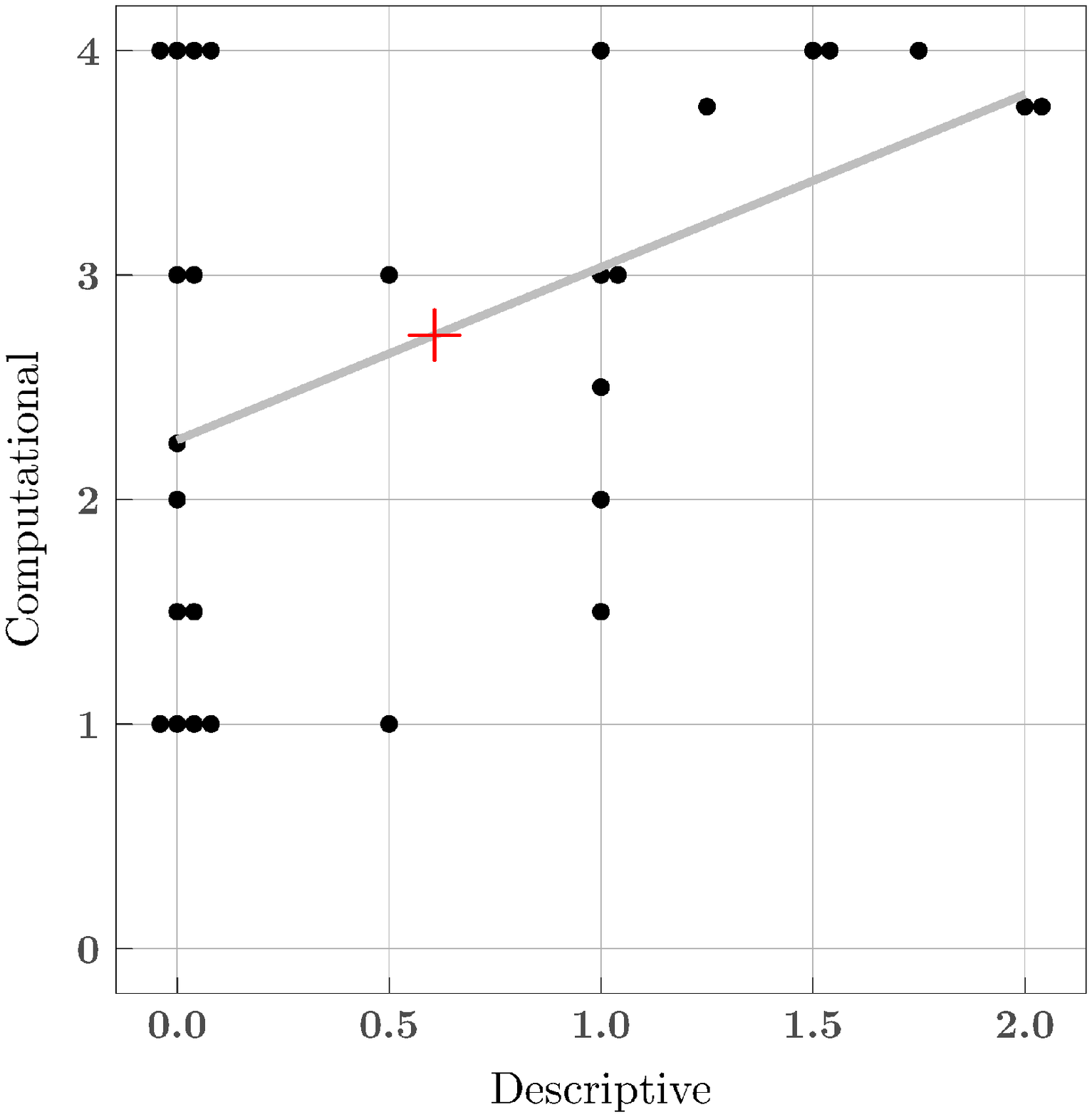}} \\
                \hline
                \parbox[c]{0.5\textwidth}{\vspace{0.5\baselineskip}\includegraphics[width=0.45\textwidth]{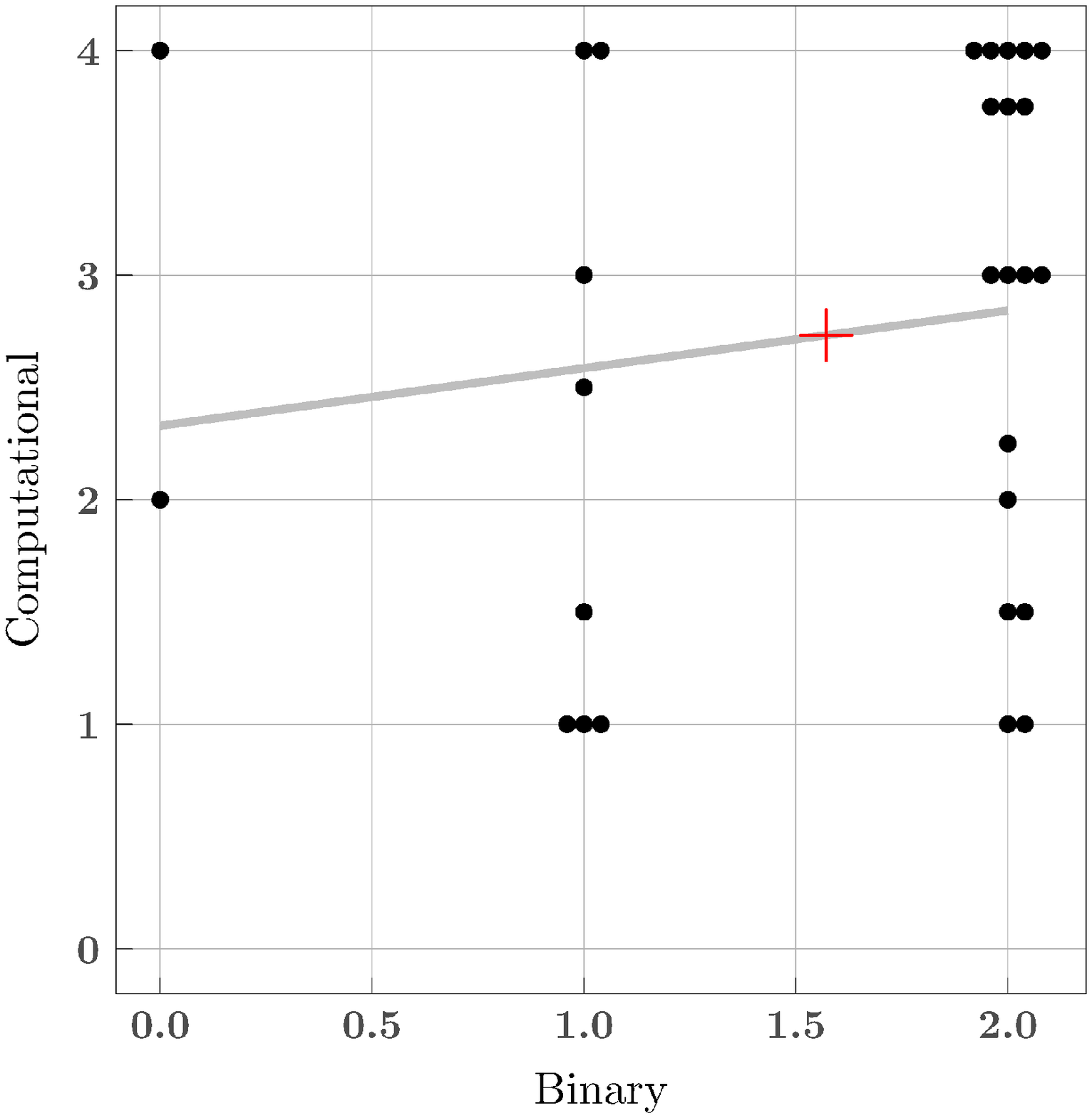}\vspace{0.5\baselineskip}} & \\
                \hline
            \end{tabular}
            \caption{\label{scatter} Correlation between responses to the binary, descriptive, and compuational questions.}
        \end{figure}

        For variables that are perfectly linearly correlated the value of $r = 1$. The calculated values suggest that the linear correlation between the three types is weak. To take a deeper look at the data we created scatter plots for the three paired combinations overlaid with the mean values and the estimated linear regression (Fig.~\ref{scatter}).

        The distributions of scores showed that there was no meaningful linear correlation between the types. The binary + computational combination in particular had a very poor correlation which suggested that student performance in one was a very poor indicator of performance in the other. We turned to simple probability to shed more light on the data.

    \subsection{Probability}

        The Tau Equivalent Reliability ($\rho_T$) and correlation coefficients revealed that the three types are likely not measuring the same construct when the entire population was considered as a whole.

        The purpose of the assessment was to determine student conceptual understanding of resistor networks. If we defined \textbf{success} in answering a question as a score revealing sufficient conceptual understanding we could use conditional probability to compare success as quantified by the three types. The threshold for defining success was objective and therefore arbitrary to some extent.

        For the binary question success was defined as a perfect score (2 points). Since only 2 students failed to get any points reducing the threshold would effectively mean that the entire class was successful, an unlikely outcome.

        In the computational and descriptive questions a threshold of 75\% of the maximum was chosen after looking at the responses in a holistic fashion. We declared B, D, and C, to be events corresponding to a student successfully answering the binary, descriptive, and computational questions respectively. Therefore to each student was assigned a True/False value based on each of these three events.

\begin{figure}[!ht]
            \includegraphics[width=0.5\textwidth]{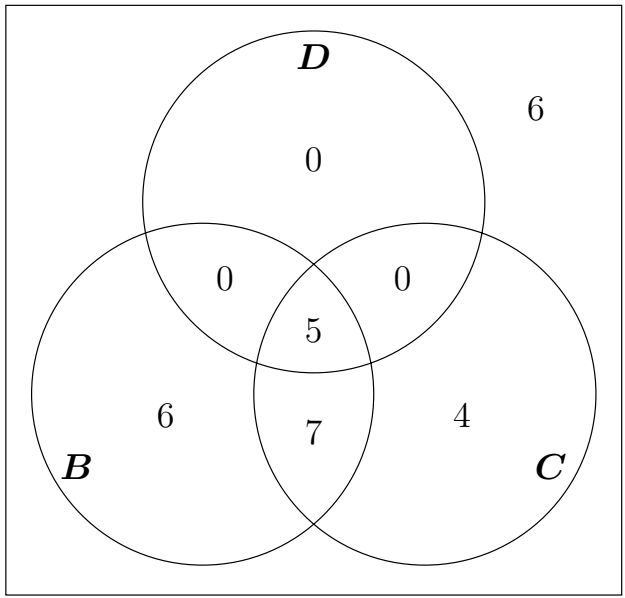}
            \caption{\label{venn} Venn Diagram of student responses}
        \end{figure}

        The students' responses were projected on to a Venn Diagram (Figure~\ref{venn}) that illustrated the distribution of events and allowed the calculation of simple, intersection, and conditional probabilities empirically from the data.\cite{conditional_prob, empirical_prob} Out of a total of 28 students, 18 got the binary question right, 16 the computational one, and only 5 were successful in answering the descriptive question. 6 students performed poorly in all three questions while 5 got all three right.

        The zeros in the D portion of the Venn Diagram led to $P(B|D) = P(C|D) = 1$, namely every student who answered the descriptive question successfully also managed to answer both the binary and computational questions correctly. However the converse was not true, $P(D|C) = 5 / 16 \approx 0.31$ and $P(D|B) = 5 / 18 \approx 0.28$ meant less than half the students who were successful in either the binary or the computational question were able to answer the descriptive question successfully. In fact $P(D|C \cap B) = 5 / 12 \approx 0.42$ told us that even amongst the students who successfully answered both the binary and computational questions less than half managed to answer the descriptive question.

        $P(D|\bar{C} \cup \bar{B}) = 0$ meant any student who failed to answer either the binary or the computational question was unable to answer the descriptive question either, while $P(B \cup C|\bar{D}) = 17 / 23 \approx 0.74$ showed that students who failed to answer the descriptive questions still stood a very good chance of getting either the binary or computational questions right.

        Another interesting tidbit hidden in the data was $P(B \cap \bar{C}) = 6 / 28$ and $P(\bar{B} \cap C) = 4 / 28$ which (being disjoint sets) meant there were 10 students whose answers to the binary and computational questions didn't agree with each other. So 35\% of the students failed to make the connection between the two parts.

\section{Judging criteria for types of questions}

    Data analysis revealed that the three types of questions were not measuring student conceptual understanding in the same way or to the same extent. However, student assessment is not the only criterion on which types of questions are judged. The following factors could be weighed by instructors before choosing a type of question for an exam: ease of design, clarity from a student perspective, ease of grading, instructiveness, and ability to assess technical skills, communication skills, and conceptual understanding.

    The questions in the case-study were designed around the computational part as the core. Once that was complete the binary and descriptive parts were added on with minimal additional effort. The descriptive part was the least clear from a student's perspective since it was open-ended and provided no directions on how a justification should be framed. This was borne out by the communicative aspect of the responses.

    The binary question was the easiest to grade. Grading the computational and descriptive parts presented distinct challenges. The computational responses were time-consuming while lack of legibility, clarity, and rigor made grading the descriptive responses difficult. In either case the time required would become prohibitive for classes with large strengths.

    An examination provides an opportunity to not only assess students but also instruct them (in the same way an assignment does). By asking the students to solve the binary part without first carrying out the detailed calculation the question attempted to open up new avenues of thought for the students. The computational part being very traditional in nature provided fewer opportunities for learning. The descriptive part on the other hand forced the students to think more deeply and more abstractly about the problem.

    The binary question by its very nature provided no opportunity to assess technical or communication skills. In fact, its ability to remove communication from the problem is one of its advantages. The computational question was an excellent means of assessing technical skills especially when the entire calculation was evaluated in detail. The descriptive question relied heavily on communication skills. The case-study however chose to overlook this aspect and focus entirely on assessing conceptual understanding.

\section{Conclusion}

    \subsection{Student Performance}

        Analyzing in detail the responses of 28 students to three types of questions (binary, descriptive, and computational) we were able to gain considerable insight in to the students conceptual understanding of resistor networks as well as their approach to solving exams.

        64.3\% of the students answered the binary question correctly. This success rate was greater than that for the computational part (57.1\%) and significantly greater than the descriptive part (17.9\%). 33.3\% of the students who got the binary question right failed to get either the descriptive or computational questions right. This was strongly indicative of guessing.

        More disturbing was the fact that as many as 10 students got just one of the binary or computational questions right. Since these questions were linked such a discrepancy should immediately have indicated to the students that they had made a mistake somewhere, either in the binary or in the computational part. These 10 students did not realize the existence of this straight-forward link. This indicates that the students were heavily compartmentalizing (mentally) while solving the question separating the first two parts from the third one.

        Since the design of the question was based on the hope that the students would use the binary part to build up an expectation for the computational part this outcome constituted sobering feedback. The insight, however, was invaluable, allowing the instructor to discuss expectation-building and continuous evaluation with the students on the basis of their responses.

        The success rate for the computational question was 57.1\% which was on the low side. This question relied on the technical skills that this course is meant to impart. Post-exam student feedback indicated that the low success rate could have been a result of the students' lack of familiarity with unseen questions (the public education system prepares them inadequately for this)\cite{christie2005rote} as well as insufficient practice solving problems.

        The disappointing number was the percentage who got the descriptive question right (17.9\%). Since the descriptive question was designed to probe the deeper (more abstract) conceptual understanding of students this result reveals severe short-comings in this regards. This ties in with general critique of Pakistan's public educational system which is often criticized for its inability to engender critical thinking in its students\cite{nauman2017lack} creating a reliance on rote memorization and general techniques for passing exams instead.\cite{christie2005rote}

        $P(B|D) = 1$ and $P(C|D) = 1$ indicated that success in the descriptive question was an absolute indicator of success in the other two questions. All 5 of the students who answered the descriptive question correctly also answered the other two questions correctly. These were obviously the better students in the class but that assessment (of who is better) was made on the basis of these questions, therefore, we can safely assert that the more abstract and deeper understanding (that is probed by the descriptive question) automatically confers fluency in the binary and computational questions.

        The perceived difficulty of such questions, their subjective nature, heavy reliance on technical communication skills and the extra effort required to grade them make descriptive questions rare in physics exams. There is then a need to change student perception about their difficulty, improve their technical communication skills (an essential requirement) and put in the extra effort to ensure that students are required to think about their work and justify their approach in a scientific manner.

        While 6 students failed to answer all three questions and the overall performance was sub-par, 5 students did manage to answer all three questions successfully. This suggests that the factors underlying the sub-optimal performance, though they might be systematic in nature, are not indicative of a lack of aptitude. The poor technical communication skills was a universal defect and stands out as an issue that requires critical attention.

    \subsection{Relative efficacy of questions}

        In addition, the data allowed us the analyze in detail the relative efficacy of the three types of questions as means of assessing student conceptual understanding of resistor networks. Before presenting the analysis we must acknowledge the small population (sample size) that answered the questions and the fact that there was only one question of each type. While this is in keeping with the nature of physics courses in general and does not detract from the study's value as an analysis of the specific student responses, it does require that any generalizations drawn be considered with a degree of skepticism.

        The Tau Equivalent Reliability parameter $\rho_T$ (Cronbach's alpha) showed that the three types were not measuring the same construct and the correlation coefficients $r$ between the types showed that the responses were poorly linearly correlated.

        In particular while comparing the binary and computational questions $\rho_T = 0.21$ and $r = 0.14$ showed, surprisingly, that the binary and computational questions were not testing the same construct at all and proficiency in one was almost completely uncorrelated with that in the other. This suggested that these two questions could not be used inter-changeably since they were testing very different things.

        To gain deeper insight in to the inter-play between the types conditional probabilities were calculated. These confirmed that the descriptive question was a strong predictor for success in all types. It was designed to assess a deeper and more abstract conceptual understanding and consequently the students found it to be the most difficult to answer.

        In light of this insight $P(B \cup C | \bar{D}) = 17 / 23 \approx 0.74$ became ominous. 74\% of the students who failed to justify their work in the descriptive question (showing a lack of deeper and abstract understanding) were still able to successfully answer the binary or computational parts.

        The binary question was designed to facilitate expectation building as a precursor to the computation. The analysis showed that it failed as a measure of conceptual understanding. Limited-choice questions are notoriously difficult to design and the time saved grading them must be balanced against the effort that is necessary to ensure they reliably assess conceptual understanding.

        For the computational question $P(\bar{D}|C) = 11 / 16 \approx 0.69$ meant 69\% of students who successfully carried out a fairly detailed computation involving the reduction of a resistor network were able to do so without demonstrating deeper conceptual understanding. They were able to combine series and parallel resistors iteratively without understanding the more abstract nature of series and parallel combinations. Had the question contained only the binary or computational parts (which is often the case in physics exams) the result would have done a disservice to both the instructor and the students by giving a false impression of the depth of the latter's conceptual understanding.

        These findings suggest that both the instructive and assessment aspects of examination are better served when limited-choice and computational questions are coupled with appropriate descriptive questions since these have the capacity to probe deeper conceptual understanding.


\begin{thebibliography}{99}
\bibitem{raw_data} Raw collected data, [URL will be inserted by AIP].
\bibitem{r_code} R scripts for analyzing and plotting case-study data, [URL will be inserted by AIP].
\end{thebibliography}
\end{document}